\documentclass[twocolumn,showpacs,aps,amsmath,amssymb,prb]{revtex4}
\usepackage{amsfonts}
\usepackage{bm}
\usepackage{graphicx}

\newcommand{\B}{\mbox{\tiny B}}

\newcommand{\ti}{\tilde}
\newcommand{\nl}{\nonumber \\}

\newcommand{\Sec}[1]{Sec.\;\ref{#1}}

\newcommand{\be}{\begin{equation}}
\newcommand{\ee}{\end{equation}}
\newcommand{\bea}{\begin{eqnarray}}
\newcommand{\eea}{\end{eqnarray}}
\newcommand{\bsube}{\begin{subequations}}
\newcommand{\esube}{\end{subequations}}

\newcommand{\Eq}[1]{Eq.\,(\ref{#1})}

\newcommand{\Fig}[1]{Fig.\,\ref{#1}}

\newcommand{\dg}{\dagger}
\newcommand{\la}{\langle}
\newcommand{\ra}{\rangle}

\newcommand{\ind}{{\sf n}}

\newcommand{\rhonup}{\rho_{\sf n}^{{ }_{\{+\}}}}
\newcommand{\rhondown}{\rho_{\sf n}^{{ }_{\{-\}}}}

\begin{document}

\title{Theoretical investigation of the
 dynamic electronic response
of a quantum dot driven by time--dependent voltage}

\author{Xiao Zheng} \email{chxzheng@ust.hk;  yyan@ust.hk}
\author{Jinshuang Jin}
\author{YiJing Yan$^{\ast}$}

\affiliation{Department of Chemistry, Hong Kong University
   of Science and Technology, Kowloon, Hong Kong}

\date{submitted on 22~July~2008s}

\begin{abstract}

 We present a comprehensive theoretical investigation on the
dynamic electronic response
 of a noninteracting quantum dot system
to various forms of time--dependent voltage applied to
the single contact lead.
Numerical simulations are carried out by implementing
a recently developed hierarchical equations of motion formalism
[J.~Chem.~Phys.~{\bf 128}, 234703 (2008)],
which is formally exact for a
fermionic system interacting with grand
canonical fermionic reservoirs, in the presence
of arbitrary time--dependent applied chemical potentials.
The dynamical characteristics of
the transient transport current evaluated
in both linear and nonlinear response regimes
are analyzed, and
the equivalent classic circuit corresponding to the coupled
dot--lead system is also discussed.

\end{abstract}


\maketitle

\section{Introduction}
\label{thintro}

Motivated by the rapid development in the field of
nanoelectronics, a comprehensive and fundamental understanding of
quantum transport phenomena has become an urging quest.
Theoretical investigations on electronic dynamics of open quantum
systems subject to external fields not only provide great insights
into relevant physical problems, but also shed light on the design
and manipulation of mesoscopic or nanoscopic electronic devices.

Frequency--dependent admittance of mesoscopic systems has been
studied by scattering theory~\cite{But93364, But934114, Pre968130}
as well as nonequilibrium Green's function (NEGF)
method.\cite{Nig06206804, Wan07155336, Ana957632} At low frequency
and low temperature, it was found that to linear order the
coherent quantum dynamics of a mesoscopic capacitor can be
characterized by a classical circuit in which a resistor and a
capacitor are connected in series.\cite{But93364, But934114,
Pre968130, Nig06206804} The resistor $R_q$, often termed as the
``charge relaxation resistor'', is related to the time scale for
electrons to accumulate on the capacitor in response to external
applied voltages, and does not rely on transmission coefficients.
For a single--channel, spin--polarized contact with a single lead,
it was predicted that at zero temperature $R_q$ is equal to half a
resistance quantum,\cite{But93364} \emph{i.e.},
\be\label{rq0}
   R_q = \frac{h}{2e^2} = 12.9 k\Omega.
\ee
The factor $1/2$ arises due to the fact that there is only one
channel connecting the quantum system to single electrode. This
has been confirmed quantitatively by a recent
experiment.\cite{Gab06499} The capacitance $C_\mu$, normally
referred to as the ``electrochemical capacitance'', was found
significantly different from the electrostatic geometric
capacitance $C_0$ due to the limited size of the quantum system
and the coherence nature of electron transport. The classical
circuit was extended afterwards to include an additional ``quantum
inductance'' $L_q$, if one of the capacitor plates is a quantum
dot (QD).\cite{Ger87271, Bro89934, Liu912705, Fu9365, Wan07155336}
$L_q$ is of purely quantum origin, and is associated with the time
scale of the resonance. For instance, consider a QD consisting of
a single spin state and coupled to one electrode. Its dynamic
admittance at zero temperature, $G(\omega)$, has been expressed
analytically by mapping to the equivalent circuit with resistance
$R_q$, capacitance $C_\mu$, and inductance $L_q$
as\cite{Wan07155336}
 \be\label{gw0}
   G(\omega) = -i\omega C_\mu + \omega^2 C_\mu^2 R_q + i\omega^3 C_\mu^3
   R_q^2 - i\omega^3 C_\mu^2 L_q.
\ee

So far, most work in this field has been conducted within the
linear response regime and based on analysis in frequency domain.
Keeping in mind that the electron transport is actually a coherent
process taking place in real time, it is thus intuitive and
straightforward to study the dynamic properties of an open quantum
system by probing its transient response to external applied
fields. The frequency--dependent admittance $G(\omega)$ can be
obtained via
\be\label{gw1}
   G(\omega) = \frac{I(\omega)}{V(\omega)} = \frac{\mathcal{F}[I(t)]}
   {\mathcal{F}[V(t)]},
\ee
where $\mathcal{F}$ denotes conventional Fourier transform. It is
worth emphasizing here that $G(\omega)$ is actually independent of
the applied voltage $V(t)$, so long as its amplitude is kept
sufficiently small. Equation~(\ref{gw1}) shows an alternative
route to evaluate $G(\omega)$ provided that the time--dependent
current response $I(t)$ can be accurately simulated via for
example quantum dissipation theory (QDT).

 We have recently constructed a formally exact QDT, in terms of
hierarchical equations of motion (HEOM), for arbitrary
non--Markovian dissipation systems interacting with Gaussian grand
canonical
ensembles.\cite{Jin07134113,Jin08234703,Xu05041103,Xu07031107} The
theoretical construction was carried out on the basis of the
calculus--on--path--integral algorithm, together with the spectral
decomposition
technique.\cite{Xu05041103,Xu07031107,Tan906676,Mei993365,Yan05187}
Dynamic responses of the reduced quantum system to external fields
can then be obtained through numerical solutions of HEOM for the
system density matrix and its associated auxiliary counterparts.
For a general many--particle system, the HEOM formalism needs to
be properly truncated at a certain finite tier, and the simulated
outcomes are considered reliable as long as they are convergent
with respect to further inclusion of higher
tiers.\cite{Jin07134113,Jin08234703,Xu05041103,Xu07031107} Great
simplification does exist for single--particle systems. It has
been proved that for the electron transport through a
noninteracting system, the HEOM formalism terminates strictly at
the second tier ($\ti{n}_{\text{max}} = 2$) without
approximation.\cite{Jin08234703}

In this work, we focus on the electron dynamics of a single--level
noninteracting QD coupled to an electrode with finite bandwidth.
The paper is organized as
follows. In \Sec{theory}, the theoretical framework and practical
implementation of the HEOM formalism is introduced. In Appendix,
numerical accuracy of the HEOM approach is validated by comparing to
exact quantum transport calculations reported in literature. In
\Sec{linear} the HEOM approach is applied to simulate the transient
electronic dynamics of an open QD. Numerical results in
linear--response regime are presented and discussed, along with a
detailed analysis of frequency--dependent dynamic admittance.
In~\Sec{nonlinear} nonlinear effects are explored. Transient current
responses to various types of applied voltages will be exemplified.
Conclusions and further comments are given in~\Sec{summary}.

\section{Methodology} \label{theory}


The QDT--HEOM formalism developed recently is formally exact
for the following standard form of Hamiltonian for quantum transport,\cite{Jin08234703}
\be \label{HT}
  H_{\rm T} = H\left(t; \{a_{\mu s}, a_{\mu s}^\dg\}\right)
    +\sum_{\alpha} (h_{\alpha}+H'_{\alpha}).
\ee
The electronic Hamiltonian ($H$)
of the system (such as QDs) is rather general, including Coulomb interaction
and  time--dependent external fields.
The electrodes are modeled by noninteracting
electrons,
\be \label{hB}
 h_{\alpha}
 = \sum_{k\in\alpha}\sum_s
   \epsilon_{\alpha ks} d^\dg_{\alpha ks} d_{\alpha ks}.
\ee
The transfer coupling $H'_{\alpha}$ between the
system and the $\alpha$--electrode reads
\be\label{Hint}
  H'_{\alpha}
=
 \sum_{k\in\alpha}\sum_{\mu s}
  t_{\alpha k\mu s}d^{\dg}_{\alpha ks}a_{\mu s}
 +{\rm H.c.}
\ee
Here, $a_{\mu s}$ $(a_{\mu s}^\dg)$ is the annihilation (creation)
operator associated with the
single--electron state $\mu$ and spin $s$ of the system,
$d_{\alpha ks}$ $(d_{\alpha ks}^\dg)$ is that
associated with the specified $\alpha$--electrode single--electron
state of energy $\epsilon_{\alpha ks}$,
and $t_{\alpha k\mu s}$ is the transfer coupling matrix element.
The transfer coupling spectral density functions are
\be\label{jpm0}
   J_{\alpha\mu\nu s}(\omega)= 2\pi \sum_{k\in\alpha}
   t^\ast_{\alpha k\mu s}\, t_{\alpha k\nu s} \,
   \delta(\omega - \epsilon_{\alpha ks}).
\ee

In contact of QDT, which describes the dynamics of reduced system
density operator $\rho(t) \equiv \mbox{tr}_{\B} \rho_{\rm T}(t)$, we
treat the electrodes as the grand canonical fermionic reservoir
bath. We denote $h_{\B}=\sum_{\alpha} h_{\alpha}$ and the
thermodynamic density operator $\rho^{\rm eq}_{\B}$ for the bare
bath in the absence of time--dependent bias voltage.
To describe the stochastic nature of the system--reservoir coupling,
consider \Eq{Hint} in the
$h_{\B}$--interaction picture,
 $H'_{\alpha}(t)
 =\sum_{\mu s}\hat f^{\dg}_{\alpha\mu s}(t)a_{\mu s}+{\rm H.c.}$,
with (setting $\hbar=1$)
\be\label{hatft}
 \hat f^{\dg}_{\alpha\mu s}(t)\equiv  e^{ih_{\B}t}
  \Big(\sum_{k\in\alpha}t_{\alpha k\mu s}d^{\dg}_{\alpha ks}
  \Big)e^{-ih_{\B}t}.
\ee
These stochastic bath operators satisfy the Gaussian statistics
with Wick's theorem for thermodynamic average over the
grand canonical fermionic bath ensembles with $\rho^{\rm eq}_{\B}$.
The effects of electrodes on the system are completely
the following nonzero reservoir correlation functions,\cite{Jin08234703}
\bsube \label{corr0}
\begin{align}
 C^+_{\alpha\mu\nu s}(t - \tau)
&\equiv
 \la \hat{f}^\dg_{\alpha\mu s}(t) \hat{f}_{\alpha\nu s}(\tau) \ra_{\B},
\\
 C^-_{\alpha\mu\nu s}(t - \tau)
&\equiv
 \la \hat{f}_{\alpha\mu s}(t) \hat{f}^\dg_{\alpha\nu s}(\tau) \ra_{\B},
\end{align}
\esube
which relate to the spectral density functions
(denoting $J^{+}_{\alpha\nu\mu s}\equiv J^{-}_{\alpha\mu\nu s}
\equiv J_{\alpha\mu\nu s}$)  via\cite{Jin08234703}
\be \label{fdt}
   C^{\pm}_{\alpha\mu\nu s}(t-\tau) = \frac{1}{2\pi}
   \int_{-\infty}^{\infty} d\omega \,
   \frac{ e^{\pm i\omega (t-\tau)}J^{\pm}_{\alpha\mu\nu s}(\omega)\,  }
   { 1 + e^{\pm\beta_\alpha\left( \omega - \mu_\alpha \right)}
   }.
\ee
Here $\beta_\alpha \equiv 1 / (k_{\B}T_\alpha)$ and $\mu_\alpha$ are
inverse temperature and equilibrium chemical potential of lead
$\alpha$, respectively.

In the presence of time--dependent voltage applied to electrodes,
the Fermi energy is subject to a homogenous time--dependent shift,
\emph{i.e.}, $\tilde{\mu}_\alpha(t) = \mu_\alpha +
\Delta_\alpha(t)$. The relevant nonequilibrium correlation functions
are\cite{Jin08234703}
\be \label{corr2}
    \tilde{C}^{\pm}_{\alpha\mu\nu s}(t, \tau) = \exp
    \left[ \pm i  \! \int_\tau^t\!\!dt' \Delta_\alpha(t')
    \right]\, C^{\pm}_{\alpha\mu\nu s}( t - \tau ).
\ee
%

From the perspective of QDT, the electronic dynamics of an open
system is mainly characterized by $\rho(t)$ and the transient
current through each lead. For linear coupling Hamiltonian of
\Eq{Hint}, system--lead dissipative interactions can be exactly
captured by the Feynman--Vernon influence functional
in path integral.\cite{Fey63118} Time derivatives on the influence functional are
performed in a hierarchical manner, and thus lead to the
construction of a formally exact hierarchical set of coupled EOM for
a general non--Markovian dissipative system.\cite{Xu05041103,%
Xu07031107, Jin07134113, Jin08234703}  The final
HEOM is cast into the following compact form
(see Ref.\,\onlinecite{Jin08234703} for details):
\be \label{eom0}
   \dot{\rho_\ind} = -\left[i\mathcal{L} + \gamma_\ind(t)  \right]
   \rho_\ind  + \rhondown + \rhonup,
\ee
with $\mathcal {L}\hat{O} \equiv [H(t), \hat{O}]$ for an arbitrary
operator $\hat{O}$.
The basic variables of \Eq{eom0} are $\rho(t)$ and associated
auxiliary density operators (ADOs) $\rho_\ind(t)$, where $\ind$ is
an index set covering all accessible derivatives of the influence
functional. With $C^\pm_{\alpha \mu \nu s}(t)$ expanded by an
exponential series via a spectral decomposition
technique,\cite{Tan906676, Mei993365, Yan05187} $\ind$ involves an
$\ti{n}$-fold combination of $(\sigma, \alpha, \mu, \nu, s, m)$ that
characterize the exponential series expansion with $\sigma=\pm$.
Therefore, $\rho_\ind \vert_{\tilde{n} = 0} \equiv \rho(t)$, and
$\gamma_\ind \vert_{\tilde{n} = 0} = \rhondown \vert_{\ti{n} = 0} =
0$; $\rho_\ind \vert_{\tilde{n} \ne 0}$ is an ADO at the
$\ti{n}^{\text{th}}$--tier; $\gamma_\ind(t)$ collects the related
exponents along with $i\Delta_\alpha(t)$, to $\rho_\ind$; and
$\rhondown$ and $\rhonup$ are the nearest lower-- and upper--tier
counterparts of $\rho_\ind$, respectively. In particular, the
$1^{\rm st}$--tier ADOs,
$\rho_\ind(t)\vert_{\tilde{n}=1}=\rho^\sigma_{\alpha\mu\nu sm}(t)$,
determine exclusively the transient current of spin--$s$ through the
lead $\alpha$ as\cite{Jin08234703}
\be \label{jts0}
   I_{\alpha s}(t)  = -2 \, \mbox{Im} \!  \sum_{\mu\nu m}
    \mbox{tr} \! \left[ a_{\mu s}\,
   \rho^+_{\alpha\mu\nu sm}(t)
   \right] .
\ee
Here the trace is performed for all QD degrees of
freedom.

The HEOM formalism has been implemented for a general dissipative
system. Details of the programming techniques will be published
elsewhere. The accuracy of our numerical approach is verified
extensively; see Appendix.

In the following we will focus on a single--level noninteracting QD
coupled to one lead, and thus omit the indexes of ($\alpha \mu \nu
s$) hereafter. The energy of the spinless QD--level is $\epsilon_0$.
An external voltage $V(t)$ is applied to the coupling lead from
$t=0$, which excites the QD out of equilibrium. The lead energy
level are shifted due to the voltage, $\Delta(t) = -e V(t)$,
and so is the lead chemical
potential $\mu(t) = \mu + \Delta(t)$, where
$e$ is the elementary charge and  $\mu$ is the equilibrium lead
Fermi energy that is set to zero hereafter, \emph{i.e.}, $\mu=0$.
A widely used Drude model is adopted for the spectral density
function of the coupling lead,
\be\label{drude0}
   J(\omega) = \frac{\Gamma W^2}
   { \omega^2 +W^2}.
\ee
In equilibrium (in absence of voltages) or in a steady state (under a
constant external voltage), the rhs of \Eq{eom0} is equal to zero,
and the HEOM reduce to a closed set of linearly coupled
equations for $\{ \rho_\ind(0)\}$. The subsequent evolution of
$\{\rho_\ind(t>0)\}$ is characterized by \Eq{eom0}, with the
equilibrium density operators as initial conditions, \emph{i.e.},
$\rho_\ind(0) = \rho^{\rm eq}_\ind$. With our current coding, the
reduced system is spanned in a Fock--state representation. The
linear sparse problem for solving the equilibrium or stead--state
$\rho_\ind$ is tackled by the biconjugate gradient
method,\cite{Pre92} and the time propagation of $\rho_\ind(t)$ in
follows the $4^{\rm th}$--order Runge--Kutta algorithm. It has been
shown\cite{Jin08234703} that for an noninteracting system, the HEOM
formalism  is {\it exact} with the terminal tier of $\ti{n}_{\rm
max} = 2$.

\section{Dynamic admittance in linear response regime} \label{linear}

\subsection{Transient electronic dynamics and frequency--dependent
 admittance}
\label{time-freq}

\begin{figure}
\includegraphics[width= 0.95\columnwidth]{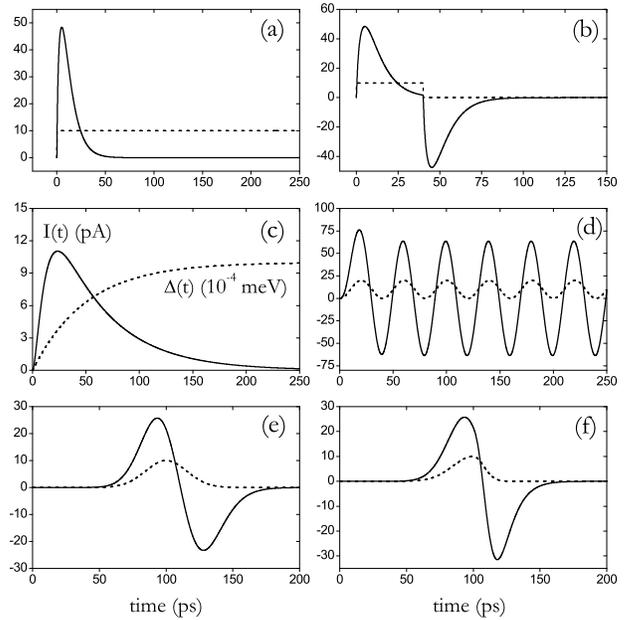}
\caption{ \label{fig1} Transient current responses to various types
of external voltages. The solid (dashed) curves represent the
time--dependent current $I(t)$ [energy shift $\Delta(t)=-eV(t)$].
The parameters adopted are as follows (in unit of meV): $\epsilon_0=
\mu = 0$, $\Gamma = 0.08$, $T = 0.026$ and $W = 3$. The switch--on
voltages are (a) a step function; (b) a square pulse; (c) an
exponential function; (d) a sinusoidal function; (e) a symmetric
Gaussian function; and (f) an asymmetric Gaussian function with the
same peak amplitude $\Delta=0.001\,$meV. }
\end{figure}

We now apply the HEOM approach to investigate the dynamic admittance
of the single--lead QD. It is presumed that the geometric
capacitance can be neglected for the QD of interest, \emph{i.e.},
$C_0 \rightarrow \infty$, which means that no charging energy is
required for an electron to populate the QD level. It is also
assumed that the QD level energy $\ti{\epsilon}_0(t) = \epsilon_0 +
\Delta_D(t)$ does not change with time, \emph{i.e.}, $\Delta_D(t) =
0$ at any time $t$, which is consistent with the $C_0 \rightarrow
\infty$ hypothesis. The dynamic admittance of a single--lead QD is
extracted from the real--time current responses $I(t)$ to external
driving voltages $V(t)$ based on \Eq{gw1}.

The general HEOM formalism admits an arbitrary form for the external
voltage $V(t)$. Transient currents due to various types of applied
$ac$ voltages are shown in \Fig{fig1}. The time--dependent
voltage adopted for \Fig{fig1}(c) corresponds to
 $\Delta(t)= \Delta (1 - e^{-t/ \tau_a})$,
 where the time constant $\tau_a > 0$
dominates the switch--on rate of the voltage, and the asymptotic
limit $\tau_a \rightarrow 0^+$ actually describes a step function.
In \Fig{fig1}(c) $\tau_a$ is taken as $50\,$ps. The driving
voltage for \Fig{fig1}(d) is a sinusoidal function: $\Delta(t) =
\Delta[1 - \cos(\omega t)]$ with the period $(2\pi/\omega) =
40\,$ps. In \Fig{fig1}(e) and (f) the open QD system is excited
out of equilibrium by a Gaussian voltage pulse as follows,
\be \label{gau1}
   \Delta(t) = \left\{
    \begin{array}{lc}
      \Delta\, e^{ - (t - \tau_c)^2 / \tau^2_b } & 0 \le t \le \tau_c \\
      \Delta\, e^{ - \kappa(t - \tau_c)^2 / \tau^2_b }& t > \tau_c
    \end{array}
    \right. ,
\ee
where $\tau_b$ and $\tau_c$ determine the width and the center of
the Gaussian pulse, respectively, and the factor $\kappa$ controls
the asymmetry of the pulse before and after $\tau_c$. In
\Fig{fig1}(e) $\tau_b = 22\,$ps, $\tau_c = 100\,$ps and $\kappa =
1$ are employed, \emph{i.e.}, the Gaussian pulse is symmetric in
time, while in \Fig{fig1}(f) $\kappa = 4$ is adopted, which means
that $\Delta(t)$ drops faster than it rises. In the linear response
regime, the dynamic admittance of the open QD system obtained by
\Eq{gw1} reflects the intrinsic physical features of the open
system, and is thus independent of the specific temporal behavior of
$\Delta(t)$, as long as its amplitude $\Delta$ is kept sufficiently
low. Among the various time--dependent voltages mentioned above, the
asymmetric Gaussian pulse is found to be a convenient candidate for
a Fourier analysis, since both $V(t)$ and the corresponding $I(t)$
have nonzero values only within a finite time interval. Therefore,
subsequent numerical analyses in this section will be based on
simulations with asymmetric Gaussian voltage pulses, if no
additional remarks are given. The purpose of introducing asymmetry
to the Gaussian voltage pulse [see \Eq{gau1}] is to save
computational effort while maintaining the accuracy of the resulting
admittance $G(\omega)$, especially in the low frequency range (to be
elaborated later). Transient current driven by some other types of
applied voltages will be discussed in~\Sec{nonlinear}, such as
step--function and delta--function voltage pulses.

It is noted that for a voltage pulse consisting of both turn--on and
turn--off sides, a sign change is always observed for the response
current~[\emph{cf.}~\Fig{fig1}(b), (e) and (f)], which shows that
an entirely positive $\Delta(t)$ can result in a negative current.
This is actually due to the fact that there is only one lead coupled
to the QD. As the voltage is turned on, the lead chemical potential
is increased, which drives the electrons flowing from the lead to
the QD (positive current) until the voltage reaches its maximum; and
then while the voltage decays, the excess electrons residing on the
QD gradually wane back to the lead, which reverses the direction of
the electron flow (negative current). As the voltage pulse vanishes
at $t \rightarrow \infty$, the initial equilibrium is restored
ultimately. Therefore, during the entire voltage on--off cycle there
is no net charge accumulating on the QD. Based on the charge
continuity equation, we should have $I(\omega=0) =
\int^\infty_{-\infty} I(t)\, dt = 0$, and this equality has been
verified for all the cases plotted in~\Fig{fig1}, except for (d)
where $\Delta(t \rightarrow \infty) \neq 0$. For the case of an
asymmetric Gaussian voltage, it is intriguing to see the negative
peak current ($31.5\,$pA) is larger than the positive counterpart
($25.8\,$pA); see~\Fig{fig1}(f). This is ascribed to the
non--adiabatic charging effect. In general, the more rapidly the
external voltage changes, the larger is the transient response
current in terms of its amplitude. Here the turn--off side of
voltage possesses a steeper slope ($\kappa > 1$), and hence the
electrons going out from the QD at $t > \tau_c$ is faster than the
rate of electron inflow at $t < \tau_c$. For the same reason, the
peak current under a step function voltage is larger than that under
an exponential function voltage;~\emph{cf}.~\Fig{fig1}(a) and
(c).

In linear--response regime, the dynamic admittance of a
noninteracting QD coupled to a single lead has been derived by the
NEGF method with the wide--band limit (WBL)
approximation.\cite{Nig06206804, Wan07155336} Extension to a finite
bandwidth case is straightforward. With a Lorentzian spectral
density function $J(\epsilon)$ [\emph{cf}.~\Eq{drude0}], the
linear--response admittance $G(\omega)$ at any temperature can be
evaluated via \Eq{gw0} for both resonant ($\epsilon_0 = \mu$) and
off--resonant ($\epsilon_0 \neq \mu$) cases with $\mu$ being the
equilibrium lead chemical potential.

\begin{figure}
\includegraphics[width= 0.95\columnwidth]{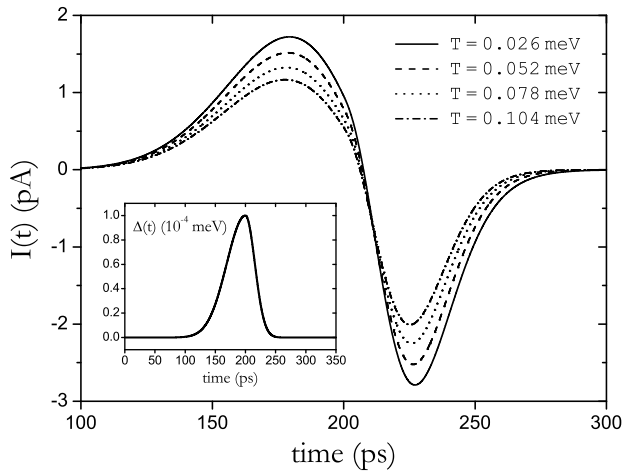}
\caption{ \label{fig2} Transient current responses to an asymmetric
Gaussian voltage pulse (see the inset) under various temperatures.
The parameters are $\kappa = 4$, and the rests in unit of
meV: $\Delta = 10^{-4}$, $\Gamma = 0.1$ and $W = 10$.
}
\end{figure}

\subsection{Resonant tunneling cases} \label{reso}

\begin{figure}
\includegraphics[width= 0.9\columnwidth]{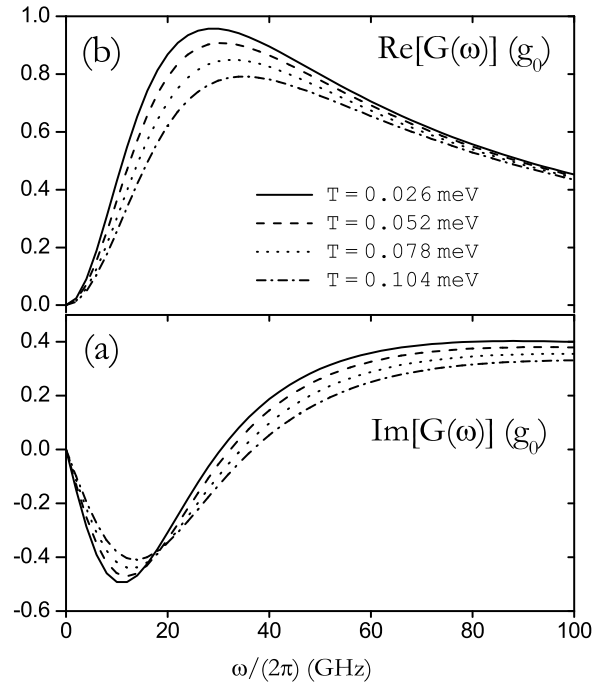}
\caption{ \label{fig3} (a) Real and (b) imaginary parts of
frequency--dependent admittance under various temperatures. Same
parameters are adopted as in~\Fig{fig2}. }
\end{figure}

We first investigate the cases where the dot level $\epsilon_0$ is
in resonance with the Fermi energy of the lead $\mu$, \emph{i.e.},
$\epsilon_0 = \mu = 0$. In~\Fig{fig2} we plot the transient
currents driven by an asymmetric Gaussian voltage pulse under
various temperatures. For all calculations carried out, the
amplitude of applied voltage is kept lower than $10\,\mu$V
throughout the simulation time to ensure the system dynamics remains
in the linear response regime. Upon the application of the driving
voltage, the QD undergoes a complete period of charge accumulation
and depletion, which is synchronized with the on--off cycle of the
external bias. It is observed that the current amplitude is
suppressed as the temperature rises. The corresponding
frequency--dependent admittances are shown in \Fig{fig3}, where
the conductance quantum $g_0 = 2e^2/h = 7.75 \times 10^{-5}\,$S is
used as the unit for $G(\omega)$. It is noted that the imaginary
part of $G(\omega)$ changes its sign within the frequency range
$\omega/ (2 \pi) \in (30, 40)\,$GHz [see~\Fig{fig3}(a)]. This
implies that the electronic dynamics of the reduced system exhibits
distinct phase features at different frequencies. In the low
frequency regime, $\mbox{Im}[G(\omega)]$ is almost proportional to
$\omega$, hence the open QD system resembles much a capacitor. Since
the geometric capacitance $C_0$ is neglected for the open QD system,
this capacitor--like behavior is of pure quantum coherence nature,
and is thus referred to as electrochemical capacitance $C_\mu$. As
$\omega$ increases, the linear relation for $\mbox{Im}[G(\omega)]$
breaks down, and the phase shift between the current and voltage
diminishes gradually until getting reversed upon the sign change.
This agrees with previous studies on the $C_\mu$ of the same open QD
system.\cite{Wan07155336}

\begin{figure}
\includegraphics[width= 0.95\columnwidth]{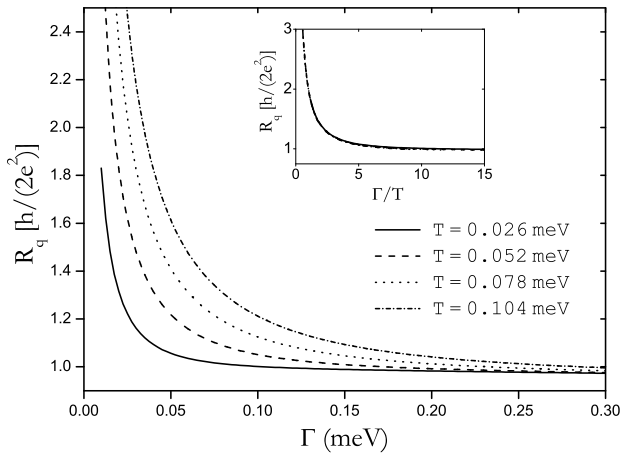}
\caption{ \label{fig4} $\mbox{Re}[Z(\omega \rightarrow 0^+)]$ as a
function of $\Gamma$ under various temperatures. Other parameters
are (in unit of meV): $\Delta = 10^{-4}$ and $W = 10$. The
inset plots $R_q$ as a function of a dimensionless quantity
$\Gamma/T$. }
\end{figure}

\begin{figure}
\includegraphics[width= 0.95\columnwidth]{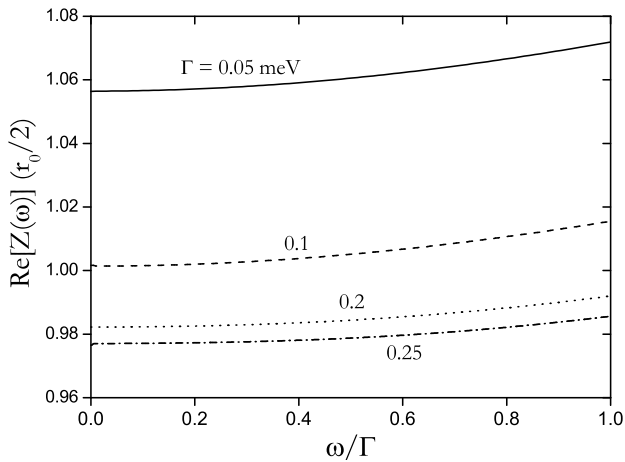}
\caption{ \label{fig5} $\mbox{Re}[Z(\omega)]$ as a function of the
dimensionless quantity $\omega / \Gamma$. The lines represent the
different values of $\Gamma$. Other parameters are (in unit
of meV): $T = 0.026$ and $W = 10$. }
\end{figure}

At zero temperature, it has been demonstrated that the dynamic
admittance of the open QD system can be quantified by a classical
$RLC$ circuit in the low frequency range ($\omega <
\Gamma$).\cite{Wan07155336} Adopting the convention in quantum
transport theory, the electrical impedance of the $RLC$ circuit is
expressed as follows [\emph{cf.} \Eq{gw0}],
\be \label{zw0}
   Z(\omega) \equiv [G(\omega)]^{-1} = R_q + \frac{1}{-i\omega C_\mu}
   - i\omega L_q.
\ee
Although $Z(\omega)$ diverges at $\omega = 0$ [due to the fact that
$I(\omega)=0$ at $\omega = 0$], $\mbox{Re}[Z(\omega)] = R_q$ is a
finite constant and independent of $\omega$, provided that $\omega
\ge 0^+$. The charge relaxation resistance $R_q$ is deduced to be
$r_0 / 2$ [\emph{cf.} \Eq{rq0}], where $r_0 = h/e^2$ is the
resistance quantum. At finite temperatures the complex impedance of
the open QD system is calculated by the HEOM approach together with
\Eq{gw1}. The resulting $\mbox{Re}[Z(\omega)]$ at $\omega
\rightarrow 0^+$ are plotted in~\Fig{fig4} as a function of
$\Gamma$. To ensure the accuracy of $Z(\omega)$, it is vital to have
high precision for the calculated $I(\omega)$, especially in the low
$\omega$ range. This can be achieved by tactically tuning the
driving voltages. For instance, introducing asymmetry to the
Gaussian pulse [\emph{i.e.}, setting $\kappa \neq 1$ in \Eq{gau1}]
magnifies the values of $I(\omega)$ as $\omega \rightarrow 0^+$, and
thus reduces its relative errors. However, no matter what type of
external voltage is applied, the HEOM approach should yield exactly
the same $G(\omega)$ in linear response regime. This has been proved
via extensive tests (see \Sec{nonlinear} for the results with
delta--function voltages).

From~\Fig{fig4} it is revealed that at a finite temperature,
$R_q$ is no longer a universal constant, but depends on the
system--bath coupling strength $\Gamma$. In particular, $R_q$
deviates significantly from the ``universal'' value $r_0/2$ when
$\Gamma$ is minute. To understand this temperature dependence of
charge relaxation resistance, we plot in the inset of~\Fig{fig4}
the calculated $R_q$ as a function of a dimensionless quantity
$\Gamma/T$ under various temperatures. All the curves are found to
overlap each other, which indicates a general trend of coherent
electronic dynamics. This thus reveals that the response current
spectrum depends parametrically on the ratio $\Gamma/T$.

As shown in the inset, $R_q$ becomes drastically larger than $r_0/2$
as $\Gamma \ll 5T$. This can be rationalized as follows. The QD
level ($\epsilon_0 = 0$) possesses an intrinsic broadening with the
magnitude $\Gamma$ due to the dissipative interaction with the lead.
At zero temperature, all the tunneling electrons are injected along
the Fermi surface of the lead ($\mu = 0$), and hence the open QD
system is in full resonance resulting in $R_q = r_0/2$. At a finite
temperature, some electrons in the lead are thermally excited.
However, as long as the energies of the excited electrons (or holes)
remain in the linewidth of the QD level,~\emph{i.e.}, $(-\Gamma,
\Gamma)$, the electronic dynamic coherence is still conserved. If
the temperature is increased further so that a significant portion
of thermally excited electrons cannot be covered by the
aforementioned energy window associated with the QD level, the
resonance is partially lost. Therefore, the mismatch between the
energy of tunneling electrons and that of QD level disfavors the
electron transport process and gives rise to a much larger $R_q$.
This is consistent with \Fig{fig5} where $\mbox{Re}[Z(\omega)]$
versus the dimensionless quantity $\omega/\Gamma$ for different
values of $\Gamma$ are plotted. It is shown that in the low
frequency ($\omega < \Gamma$) regime, $\mbox{Re}[Z(\omega)]$ is
rather insensitive to $\omega$ (with the maximal deviation of
roughly $3\%$ at $\omega = \Gamma$) for all simulated cases.
However, $\mbox{Re}[Z(\omega)]$ assumes a larger average value as
$\Gamma$ becomes weaker. Therefore, we conclude that at a finite
temperature, the charge relaxation resistance $R_q$ is no longer a
universal constant, but depends on both the temperature and the
coupling strength $\Gamma$.

\subsection{Off--resonant tunneling cases} \label{off-reso}

\begin{figure}
\includegraphics[width= 0.95\columnwidth]{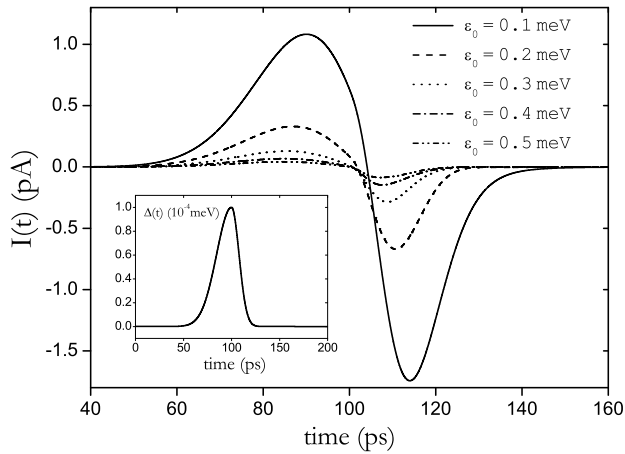}
\caption{ \label{fig6} Transient current responses to an asymmetric
Gaussian voltage pulse (see the inset) for various off--resonant
cases. The parameters are $\kappa = 4$, $\tau_c = 100\,$ps,
$\tau_b = 22\,$ps, and the rests in unit of meV: $\Delta = 10^{-4}$,
$T = 0.104$, $\Gamma = 0.1$ and $W = 10$.}
\end{figure}

\begin{figure}
\includegraphics[width= 0.95\columnwidth]{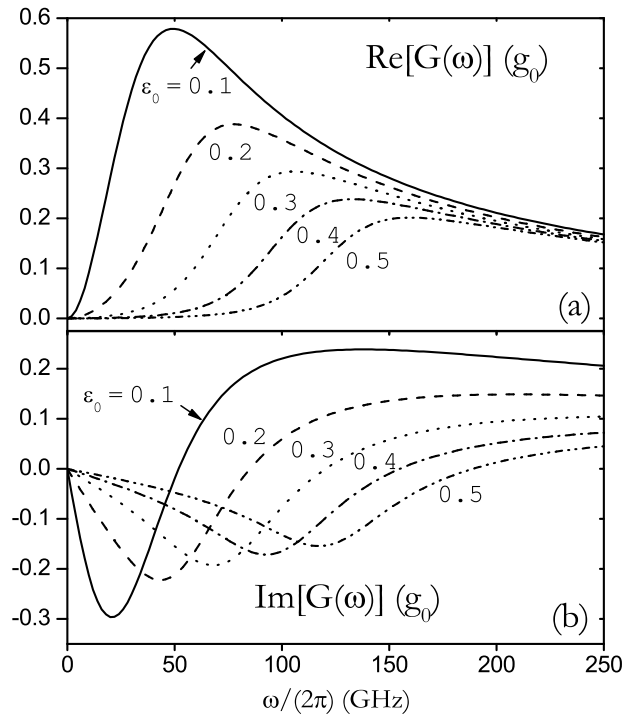}
\caption{ \label{fig7} (a) Real and (b) imaginary parts of
$G(\omega)$ for various off--resonant cases. Same parameters are
adopted as in~\Fig{fig6}. In both (a) and (b), the lines represent
different $\epsilon_0$ in unit of meV.}
\end{figure}

We now turn to the cases where $\epsilon_0 \neq \mu$. It is
important to point out that as $\epsilon_0$ rises away from $\mu$, a
much larger number of exponential functions needed to be adopted for
the expansion of $C^\pm_{\alpha\mu\nu s}(t)$ to guarantee the
accuracy of the outcomes. It is inferred that the short--memory
components of bath correlation functions play significant roles in
the off--resonant cases. Figure~\ref{fig6} depicts the transient
current responses to an asymmetric Gaussian voltage pulse for
various dot energies $\epsilon_0$. The current profiles are
analogous to the resonant cases shown in \Fig{fig2}, except that
the peak amplitude of $I(t)$ declines drastically as $\epsilon_0$
deviates continually from $\mu$. The corresponding
frequency--dependent admittances are depicted in~\Fig{fig7},
where both the real and imaginary parts of $G(\omega)$ exhibit
conspicuous blue shifts with the increasing $\epsilon_0$.
Intuitively, as the dot level goes up, fewer electrons can tunnel
through the potential barrier at the system--bath interface and
dwell on the QD, and hence the electrochemical capacitance of the
reduced system becomes smaller. This coincides with the tendency
shown in~\Fig{fig7}(a), where the descending slope of
$\mbox{Im}[G(\omega)]$ in the low $\omega$ range becomes less steep
as $\epsilon_0$ increases.

\section{Transient electronic dynamics in nonlinear response regime}
\label{nonlinear}

As the applied voltage intensifies, the dynamic response of the open
QD system goes beyond the linear response regime. In such a case,
the associated frequency--dependent admittance depends explicitly on
the specific type of the external voltage. It is thus difficult to
derive a general expression for $G(\omega)$, and the electronic
dynamics needs to be studied case by case.

Since the HEOM approach admits arbitrary time--dependent applied
voltages, no extra effort is required for calculations under higher
external biases,~\emph{i.e.}, the numerical procedures of \Eq{gw1}
can be extended straightforwardly to the nonlinear response regime.
In this section, transient currents driven by three types of
turn--on voltages will be presented in the subsections: (a) an
asymmetric Gaussian function, (b) a step function, and (c) a delta
function.

\subsection{Frequency--dependent admittance under asymmetric
Gaussian voltage pulses} \label{gw_agauss_nonlinear}

\begin{figure}
\includegraphics[width= 0.95\columnwidth]{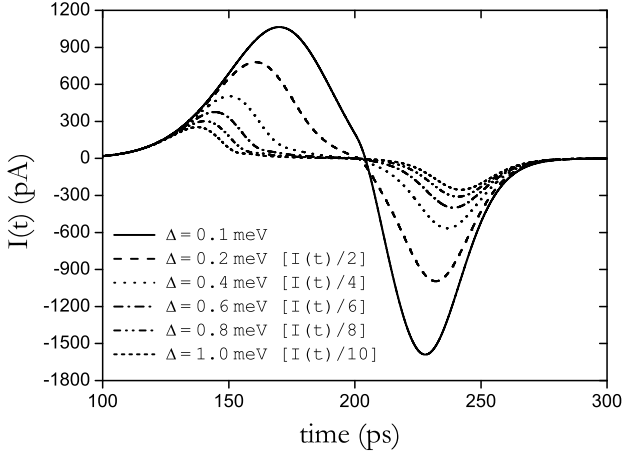}
\caption{ \label{fig8} Scaled transient currents under asymmetric
Gaussian voltages of different amplitudes. The parameters
are $\kappa = 4$, $\tau_c = 200\,$ps, $\tau_b = 44\,$ps, and the
rests in unit of meV: $T = 0.078$, $\Gamma = 0.1$ and $W = 5$.}
\end{figure}

\begin{figure}
\includegraphics[width= 0.95\columnwidth]{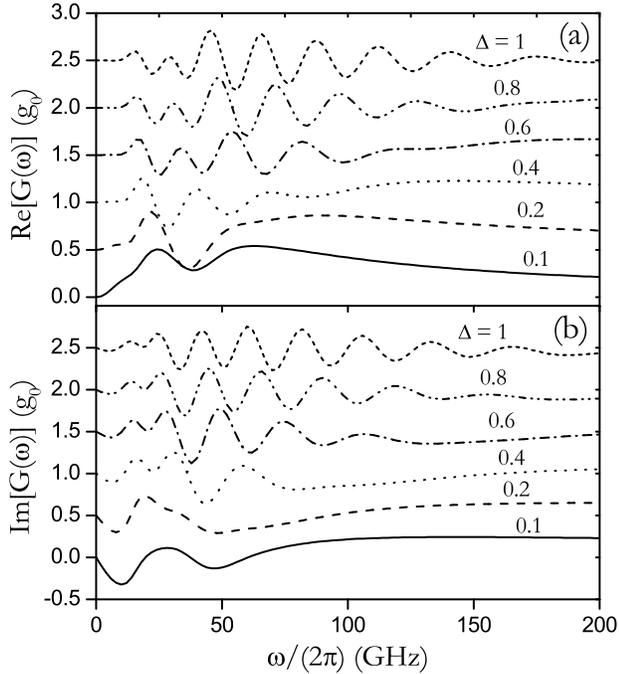}
\caption{ \label{fig9} (a) Real and (b) imaginary parts of
$G(\omega)$ under Gaussian voltages of different amplitudes. For
both (a) and (b), the lines represent different $\Delta$ in unit of
meV. Same parameters are adopted as in \Fig{fig8}. In both panels
the lines are separated vertically by $0.5\,g_0$ for clarity. }
\end{figure}

In \Fig{fig8} we plot the transient currents under asymmetric
Gaussian voltages of the amplitudes $\Delta$ ranging from $0.1$ to
$1\,$meV (recall that in linear response cases, $\Delta$ is set
lower than $10^{-4}\,$meV). It is observed that as $\Delta$
increases, the electron accumulation and depletion periods become
separated from each other. Especially for $\Delta = 1\,$meV, the
transient current almost vanishes in the time interval of $150 \sim
200\,$ps, whereas in the due course $\Delta(t)$ still keeps rising.
The corresponding frequency--dependent admittances are shown in
\Fig{fig9}, where two adjacent lines are separated vertically by
$0.5\,g_0$ for clarity. It is evidently indicated that as the
applied voltage increases, higher--energy current components are
gradually activated, and the resulting $G(\omega)$ appears more
fluctuating. The complicated lineshape of $G(\omega)$ seems to
exclude any simple equivalent classical circuit that can describe
quantitatively the coherent electronic dynamics of the open QD
system.

\subsection{Transient current driven by a step function voltage and
its spectrum analysis} \label{jw_step}

The exact time--dependent current $I(t)$ driven by a step function
voltage has been obtained by the NEGF method.\cite{Mac06085324}
Especially with the WBL approximation ($W \rightarrow \infty$), its
Fourier transform $I(\omega)$ can be evaluated conveniently via an
EOM for the reduced single--electron density matrix for the reduced
system.\cite{Zhe07195127} Under a step function voltage, the lead
level shift is $\Delta(t) = \Delta \Theta(t)$. With a tiny $\Delta$,
the linear--response admittance as well as the associated charge
relaxation resistance $R_q$ should be reproduced from the above
derivations for $I(\omega)$.\cite{Mo08Paper1}

\begin{figure}
\includegraphics[width= 0.95\columnwidth]{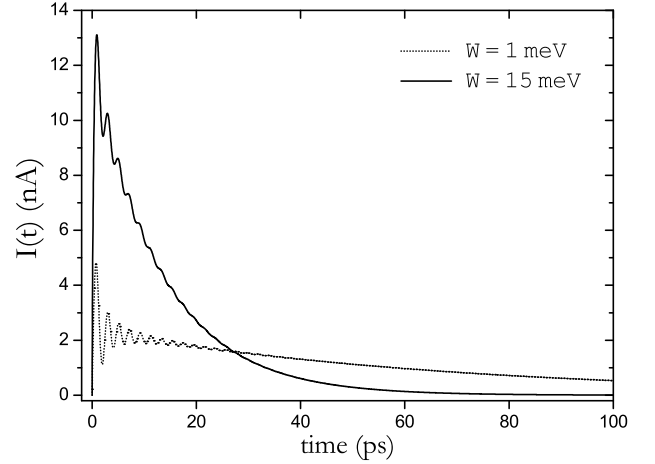}
\caption{ \label{fig10} Transient current responses to a step
function voltage. The lines represent different bandwidths. Other
parameters (in unit of meV): $\Delta = 5$, $T = 0.02$,
$\Gamma = 0.05$ and $\epsilon_0 = 3$.}
\end{figure}

\begin{figure}
\includegraphics[width= 0.95\columnwidth]{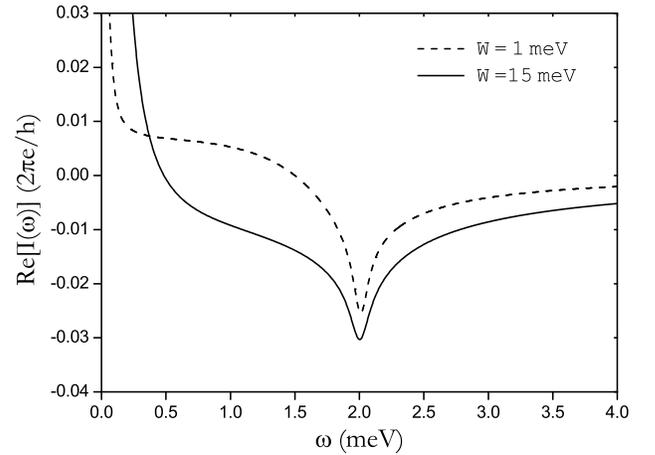}
\caption{ \label{fig11} The real part of $I(\omega)$. The inset
magnifies the lines around the valleys. Results with finite $W$
correspond to $I(t)$ in \Fig{fig10}. The other parameters are (in
unit of meV): $\Delta = 5$, $\Gamma = 0.05$ and $\epsilon_0 = 3$. }
\end{figure}

\begin{figure}
\includegraphics[width= 0.95\columnwidth]{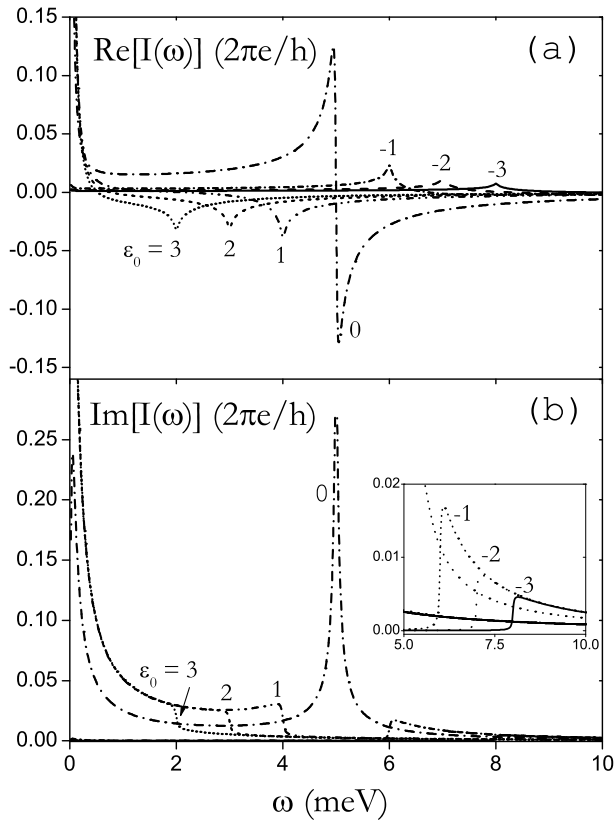}
\caption{ \label{fig12} (a) Real and (b) imaginary parts of
$I(\omega)$ under a step function voltage of $\Delta = 5\,$meV. The
lines represent different $\epsilon_0$ in unit of meV. The inset
magnifies the down--right corner of panel (b). The other parameters
are (in unit of meV): $T = 0$, $W = \infty$ and $\Gamma = 0.05$.}
\end{figure}

The transient currents calculated by the HEOM approach are plotted
in \Fig{fig10}. It is shown that with a narrower lead conduction
band, the initial overshooting of $I(t)$ becomes less prominent and
it takes longer time for the open system to reach the steady state.
High--frequency oscillations are clearly observed for $I(t)$ at the
time $0 \le t \le 25\,$ps, which are due to the large amplitude of
applied voltage. This nonlinear effect is confirmed by making a
comparison to \Fig{fig1}(a), where the rapid oscillation is
absent from the transient current due to the much smaller $\Delta$.
It is revealed from the Fourier analysis that the characteristic
oscillation frequency is centered at $\omega_0 = \Delta -
\epsilon_0$, as depicted in \Fig{fig11}. In \Fig{fig12} we further
investigate $I(\omega)$ by plotting its dependence on $\epsilon_0$
under a fixed $\Delta$. For all cases $I(\omega)$ either reaches an
extreme point or undergoes a sudden change in terms of its value at
$\omega = \omega_0$. It is interesting to note that for $\epsilon_0
< \mu$ the plotted $\mbox{Re}[I(\omega)]$ exhibits a peak at $\omega
= \omega_0$, while for $\epsilon_0 > \mu$ a dip shows up. As
$\epsilon_0$ is drawn closer to $\mu$, the current response at the
frequency $\omega_0$ is more accentuated. In the time domain, this
implies an enhanced oscillation amplitude for the transient current.
For the resonant case, \emph{i.e.}, $\epsilon_0 = \mu$, the function
form of $\mbox{Re}[I(\omega)]$ near $\omega_0$ is $\ln(x^2 + 1)/x$
with $x = 2(\omega - \omega_0)/\Gamma$, which is different from the
typical Fano line shape $(x + q)^2 / (x^2 + 1)$.\cite{Fan611866}

\begin{figure}
\includegraphics[width= 0.95\columnwidth]{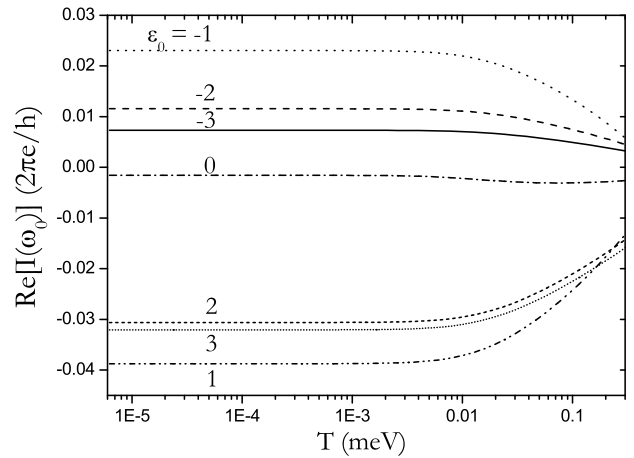}
\caption{ \label{fig13} $\mbox{Re}[I(\omega_0)]$ as a function of
temperature. The lines represent different values of $\epsilon_0$ in
unit of meV. The horizontal axis is in $\log_{10}$ scale. Other
parameters are: $W = \infty$ and $\Gamma = 0.05\,$meV. }
\end{figure}

The thermal influence on the transient current is also explored. In
\Fig{fig13} we plot $\mbox{Re}[I(\omega_0)]$ as a function of
temperature. It is found that the system under investigation starts
to respond sensitively to the environmental temperature at $T \sim
0.01\,$meV. Upon further increase of $T$, the thermal effect
overwhelms the system--bath--coupling--induced linewidth, which is
$0.05\,$meV in this specific case, and thus dominates the electronic
dynamics of the reduced system.

\subsection{Transient current driven by a delta function voltage and
its spectrum analysis} \label{jw_delta}

A delta function external field has been used to study the transient
electronic dynamics of an isolated molecular
system.\cite{Wan07134104} For an open system, such as the
single--level QD system of our primary interest, a delta function
applied voltage is also useful to investigate its dynamic
properties. The time--dependent level shift considered is $\Delta(t)
= \Delta \Theta(t) \delta(t)$. The advantage of such a delta
function voltage is two--fold: (1) its Fourier transform
$\Delta(\omega) = \Delta / 2$ is a constant, thus the current
response of any frequency can be detected, with $I(\omega) \propto
G(\omega)$; (2) the HEOM~(\ref{eom0}) becomes a set of
time--independent linear equations at $t \ge 0^+$, and an efficient
Chebyshev propagator can be employed to solve the evolution of
$\rho_\ind(t)$,\cite{Bae049803, Wan07134104} which greatly reduces
the computational cost.

To simulate the transient current in response to a delta function
voltage by the HEOM approach, we notice that $\Delta(t)$ takes
effect only within the infinitesimal interval, \emph{i.e.}, $t \in
(0, 0^+)$. Therefore, the reduced dynamics at ($t > 0^+$) can be
solved by normal propagation of \Eq{eom0} (in absence of bias
voltages) with the initial condition adjusted to $\rho_\ind(0^+)$.
Consider a moderate amplitude $\Delta$, so that all the ADOs remain
finite within $t \in (0, 0^+)$. Therefore, neither $\rhonup$ nor
$\rhondown$ contributes as \Eq{eom0} is formally integrated from
time $t = 0$ to $0^+$:
\begin{align} \label{rho0p}
   \rho_\ind(0^+) &= \rho_\ind(0) + i \sum_{\ind} \, \sigma_\ind
   \int_0^{0^+} \!\! \Delta(\tau) \rho_\ind(\tau) d\tau
   \nl&= \frac{1}{1 - i\ti{\Delta}_\ind}\, \rho_\ind(0).
\end{align}
Here $\rho_\ind(0)$ are the equilibrium reduced density matrix and
associated ADOs, and $\ti{\Delta}_n$ is expressed as follows,
\be \label{tidelta}
  \ti{\Delta}_\ind \equiv \sum_{\ind} \sigma_\ind \int_0^{0^+}\!\!
  \Delta(\tau) d\tau = \frac{\Delta}{2} \sum_{\ind} \sigma_\ind.
\ee
Due to the fact that $\ti{\Delta}_\ind \vert_{\ti{n}=0} = 0$ we have
$\rho(0^+) = \rho(0)$, \emph{i.e.}, the reduced density matrix
remains continuous upon the delta function perturbation.

\begin{figure}
\includegraphics[width= 0.95\columnwidth]{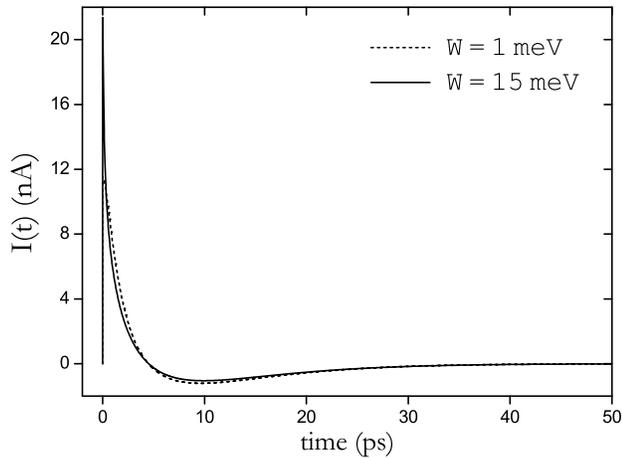}
\caption{ \label{fig14} Transient current responses to a delta
function voltage. The lines represent different bandwidths.
Other parameters are (in unit of meV): $\Delta = 1$, $T = 0.052$,
$\Gamma = 0.1$ and $\epsilon_0 = 0$.}
\end{figure}

\begin{figure}
\includegraphics[width= 0.95\columnwidth]{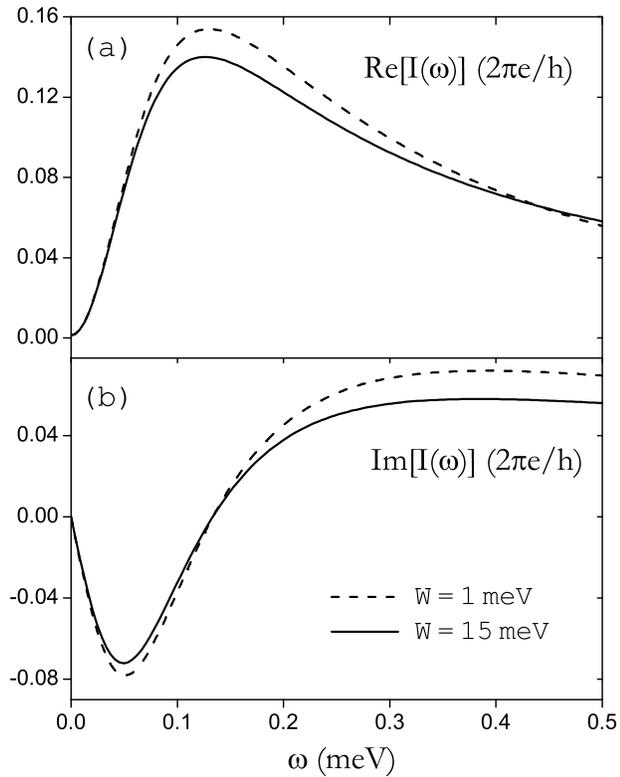}
\caption{ \label{fig15} (a) Real and (b) imaginary parts of
$I(\omega)$ under a delta function voltage. Results with finite $W$
correspond to $I(t)$ in \Fig{fig14}. The common parameters are (in
unit of meV): $\Delta = 1$, $\Gamma = 0.1$ and $\epsilon_0 = 0$. }
\end{figure}

In \Fig{fig14} we plot the calculated transient currents
corresponding to different bandwidths. In both cases the current is
instantaneously switched on to its maximal value at $t = 0^+$, and
then relaxes back to zero. The peak value $I_{\text max} = I(0^+)$
is $21.4\,$nA for $W = 15\,$meV, and $11.5\,$nA for $W = 1\,$meV,
respectively. The corresponding current spectrums are depicted in
\Fig{fig15}. The similarity in lineshape between \Fig{fig3} and
\Fig{fig11} is noted, although the former is a linear response
result while the latter belongs to the nonlinear regime. This is
actually due to the factorization property of $I(\omega)$ for the
resonant tunneling case,\cite{Mo08Paper1} where the current spectrum
can be expressed as $I(\omega) = \sin(\Delta/2) X(\omega)$ in the
resonant case ($\epsilon_0 = \mu = 0$), where $X(\omega)$ is some
complex function independent of $\Delta$.

\section{Conclusions and comments} \label{summary}

To conclude, we investigate the quantum coherent electronic dynamics
of a single--level noninteracting QD coupled to one electrode.
Simulations are carried out based on the HEOM formalism of
QDT.\cite{Jin08234703} In the linear response regime, quantitatively
accurate frequency--dependent admittance is obtained by the
calculated transient current response to an asymmetric Gaussian
voltage pulse, for both resonant and off--resonant tunneling cases.
It is verified that at a finite temperature, the dynamic admittance
of the open QD system in the low frequency range can still be
characterized by a classical $RLC$ circuit. However, the charge
relaxation resistance $R_q$ is found deviated from half a resistance
quantum, but depends on both the temperature and the system--bath
coupling strength. The concept of the equivalent classical circuit
breaks down under higher bias. Complicated nonlinear features
are observed, such as the activation of high--frequency current
components with an increasing voltage amplitude. Transient current
responses to step and delta function voltages are also explored. The
basic features of the associated electronic dynamics are analyzed
and discussed. The analytical and numerical results obtained in
Sec.~\ref{linear} and~\ref{nonlinear} serve as convenient basis to
understand the electronic dynamics of an interacting open system,
in which the HEOM formalism is in principle exact.
Work along this direction is underway.

The HEOM approach also has great implications for time--dependent
quantum transport in nanoscopic molecular devices. Formalisms based
on NEGF method,\cite{Che903159,Lan912541,Afo9410466,Jau945528,%
Lop01075319,Kne03066122,Zhu05075317,Kur05035308,Qia06035408,%
Mold07165308,Mol07085330,Sou07125318,Wei08195316}
Floquet theory,\cite{Sta961916,Tik0210909,Cre02113304,Cre02235303,%
Pla041,Bra04205326}
and QDT~\cite{Cui06449,Li07075114,Wel06044712,Vil08192102}
have been proposed. First--principles
calculations have been carried out on realistic electronic
devices.\cite{Zhe07195127} However, in these simulations,
approximate schemes for the dissipative dynamics in real time
are inevitably introduced. For instance, the
complete second--order formulation~\cite{Yan982721, Xu029196,
Yan05187} and the WBL approximation were adopted to simulate the
steady and transient current through a molecular device with
time--dependent density--functional theory.\cite{Zhe07195127} It is
thus important that these approximate schemes can be improved
systematically. Since the QDT--HEOM approach is capable of yielding
exact results for quantum dissipative dynamics, it can be utilized
to calibrate the approximated methodologies, and we also expect it
to provide some guidelines for the potential progress of the
approximate formalisms.

\begin{acknowledgments}
   Support from the RGC (604007 and 604508) of Hong Kong is acknowledged.
\end{acknowledgments}

\appendix*
\section{Numerical validation}
\label{test-td}

We verify our numerical implementation of HEOM formalism by comparisons to
known exact quantum transport results achievable via other methods.
Three cases of noninteracting QDs coupled to
left and right electrodes ($L$ and $R$) of
Lorentzian spectral density functions are demonstrated as follows.

\begin{figure}
\includegraphics[width=0.95\columnwidth]{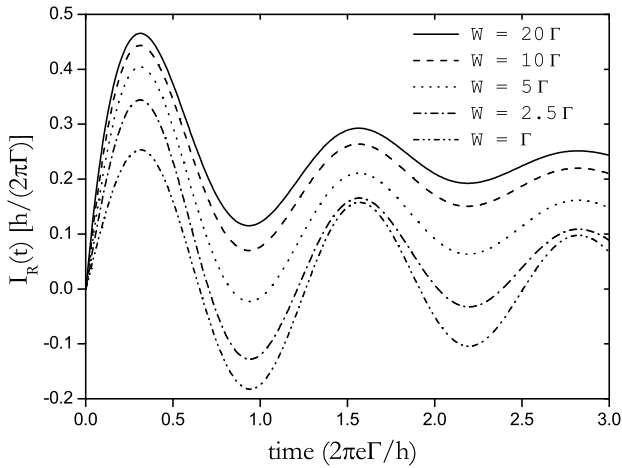}
\caption{ \label{fig16} Transient current through R--lead,
$I_{R}(t)$, in response to a step--function voltage pulse applied on
R--lead. The lines correspond to different lead bandwidths. Other
parameters: $\Gamma_L = \Gamma_R
= 0.5\,\Gamma$, $T= 0.1\,\Gamma$, $W_L = W_R =
W$ and $\Delta = 10\,\Gamma$.
This figure reproduces Fig.\,2 in Ref.~\onlinecite{Mac06085324}.}
\end{figure}

The first case studies the time--dependent transport through a
single--level spinless QD driven by a step--function voltage pulse,
calculated  before exactly by the NEGF
method.\cite{Mac06085324, Ped05195330, Ste04195318}
In \Fig{fig16} we plot the transient currents calculated via the HEOM approach
on the same system as that of Fig.\,2 in Ref.\,\onlinecite{Mac06085324}.
Our results agree quantitatively.
The QD is initially in equilibrium with zero bias. External voltage pulses are
switched on from the time $t = 0$, which results in current flows
through the leads $L$ and $R$ at $t > 0$. Hereafter we denote
$\Delta_L(t)$ and $\Delta_R(t)$ as the time--dependent energy shifts
for the lead $L$ and $R$ due to the applied voltages $V_L(t)$ and
$V_R(t)$, respectively, \emph{i.e.}, $\Delta_\alpha(t) = -e
V_\alpha(t)$. For simulations presented in this section, we set
$\Delta_L(t) = 0$ for all $t$, and $\Delta_R(t) = \Delta \Theta(t)$,
where $\Theta(t)$ is a step function turned on at $t = 0$, and
$\Delta$ is a constant amplitude.
The system level energy is time--dependent as $\ti{\epsilon}_0(t) = \epsilon_0 +
\Delta_D(t)$ with $\Delta_D(t) = [\Delta_L(t) + \Delta_R(t)]/2$.

\begin{figure}
\includegraphics[width= 0.95\columnwidth]{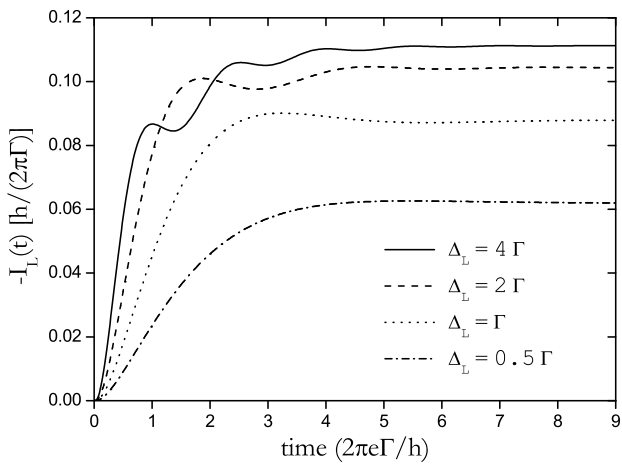}
\caption{ \label{fig17}
  Transient currents through L--lead,
$I_L(t)$, in response to a step--function voltage pulse applied on
R--lead. The lines correspond to different voltage amplitudes.
Other parameters: $\Gamma_L = \Gamma_R
= 0.5\,\Gamma$, $T = 0.05\,\Gamma$ and $W_L = W_R =
20\,\Gamma$.
This figure reproduces Fig.\,1 in Ref.~\onlinecite{Ped05195330}.
}
\end{figure}

\begin{figure}
\includegraphics[width= 0.95\columnwidth]{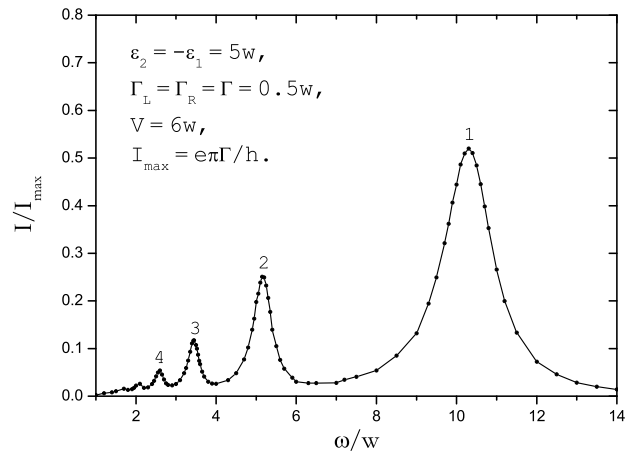}
\caption{ \label{fig18}
  Time--averaged current $I$ (in unit of
$I_{\rm max} = e\Gamma/2\hbar$) through a noninteracting double QD,
driven by {\it ac} gate voltage
(see Ref.~\onlinecite{Sta961916} for details).
The interdot coupling strength $w = 0.1\,$meV.
Other parameters: $\mu^{\rm eq}_L = \mu^{\rm eq}_R = 0$,
$T = w$ and $W_L = W_R = 20\,w$.
The characteristic multiphoton--assisted resonance energies
$\omega_N = \frac{1}{N}\sqrt{(\epsilon_2 -\epsilon_1)^2 + 4w^2}$ are
labeled by numbers
$N=1 \sim 4$. This figure reproduces Fig.\,3 in
Ref.~\onlinecite{Sta961916}. }
\end{figure}

\begin{figure}
\includegraphics[width= 0.95\columnwidth]{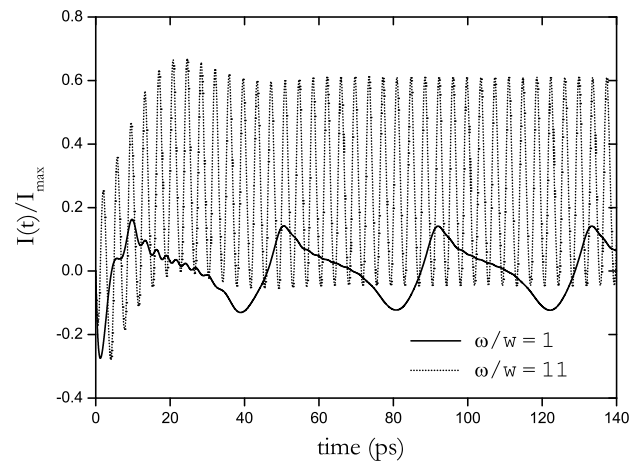}
\caption{ \label{fig19} Transient current $I_L(t)$ (in unit of
$I_{\rm max} = e\Gamma/2\hbar$) through a noninteracting double QD.
The system setup and other parameters are same as in \Fig{fig18}.
}
\end{figure}

The second example is similar to the first one, but with
$\Delta_D(t) = 0$. The calculated transient currents displayed in \Fig{fig17}
accurately reproduce previous simulations
(\emph{cf}.~Fig.\,1 in Ref.\,\onlinecite{Ped05195330}) by the NEGF
method with all the non--Markovian features
preserved.\cite{Ped05195330,Ste04195318}

The third example investigates the resonant photon--assisted tunneling through QDs
where results have been obtained by a combined method of NEGF and Floquet
formalisms.\cite{Sta961916,Pla041} In Ref.\,\onlinecite{Sta961916},
nonequilibrium electron pumping through a double QD system driven by
$ac$ gate voltage was simulated. Multi--photon--assisted tunneling
was resolved in time--average current spectrum. Since the HEOM
approach admits an arbitrary time--dependent external field, a
sinusoidal gate voltage can be treated readily. However, instead of
the frequency--domain calculation conducted in
Ref.\,\onlinecite{Sta961916}, with our present coding scheme we need
to propagate \Eq{eom0} in real time for every individual frequency.
This makes the simulation for the entire current spectrum rather
tedious. Nonetheless, to further justify our numerical procedures,
we have managed to carry out such a simulation and compared to
reported results. A noninteracting double QD system coupled to two
leads is studied. The time--averaged current is evaluated as the
system reaches a quasi--steady state for each gate voltage frequency
$\omega$. We take the same parameter set as adopted by Fig.\,3 in
Ref.\,\onlinecite{Sta961916}, and the HEOM calculation result is
plotted in \Fig{fig18}. The quantitative agreement between our
\Fig{fig18} and result shown in Ref.\,\onlinecite{Sta961916} is
noted. In particular, the characteristic $N$--photon--assisted
resonance frequencies are correctly reproduced, as demonstrated
in \Fig{fig18}. This test again validates our numerical
procedures.
It is worth mentioning that the Floquet formalism treats the
quasi--steady dynamics driven by an $ac$ external field, while our
HEOM formalism allows access to much broader information beyond
this, such as the establishment of a quasi--steady state in real
time. For instance, \Fig{fig19}(a) and (b) depict the time evolution
of a double QD driven by an $ac$ gate voltage switched on from $t=0$
for $\omega/w=0.1$ and $1.1$, respectively, where $w$ is the
inter--dot coupling strength as the system setup is the same as in
\Fig{fig18}.

%


%
%

\end{document}